\documentclass[IEEEtran]{article}
\usepackage{spconf}
\usepackage{graphicx, amsfonts}
\usepackage{amsmath, amssymb}
\usepackage{cite, array, multirow}
\usepackage{siunitx}
\usepackage{xcolor, color, hhline, stfloats}
\usepackage{lipsum}
\usepackage{gensymb}
\usepackage{verbatim,setspace}
\renewcommand{\baselinestretch}{0.95}

\title{Enhancing END-TO-END MULTI-CHANNEL SPEECH SEPARATION via spatial feature learning}
\name{Rongzhi Gu$^{1}$ \thanks{Rongzhi Gu performed this work during internship at Tencent AI Lab. This paper was partially supported by Shenzhen Science \& Technology Fundamental Research Programs (No:JCYJ20170817160058246 \& JCYJ20180507182908274). $\star$ zouyx@pku.edu.cn} \qquad Shi-Xiong Zhang$^{2}$ \qquad Lianwu Chen$^{3}$ \qquad Yong Xu$^{2}$ \qquad Meng Yu$^{2}$ \\ Dan Su$^{3}$ \qquad Yuexian Zou$^{1,4\star}$ \qquad Dong Yu$^{2}$
}

\address{$^{1}$ ADSPLAB, School of ECE, Peking University, Shenzhen, China \\
$^{2}$Tencent AI Lab, Bellevue, WA, USA \\
$^{3}$Tencent AI Lab, Shenzhen, China \\
$^{4}$Peng Cheng Laboratory, Shenzhen, China}
		
\begin{document}
%
\maketitle
\begin{abstract}
Hand-crafted spatial features (e.g., inter-channel phase difference, IPD) play a fundamental role in recent deep learning based multi-channel speech separation (MCSS) methods. 
However, these manually designed spatial features are hard to incorporate into the end-to-end optimized MCSS framework.
In this work, we propose an integrated architecture for learning spatial features directly from the multi-channel speech waveforms within an end-to-end speech separation framework. In this architecture, time-domain filters spanning signal channels are trained to perform adaptive spatial filtering. These filters are implemented by a 2d convolution (conv2d) layer and their parameters are optimized using a speech separation objective function in a purely data-driven fashion. Furthermore, inspired by the IPD formulation, we design a conv2d kernel to compute the inter-channel convolution differences (ICDs), which are expected to provide the spatial cues that help to distinguish the directional sources. Evaluation results on simulated multi-channel reverberant WSJ0 2-mix dataset demonstrate that our proposed ICD based MCSS model improves the overall signal-to-distortion ratio by 10.4\% over the IPD based MCSS model.


\end{abstract}
\section{Introduction}
\label{sec:intro}

Speech separation refers to recovering the voice of each speaker from overlapped speech mixture. It is also known as cocktail party problem \cite{Cherry1960Contribution}, which has been studied in signal processing literature for decades. 
Leveraging the power of deep learning, many methods have been proposed for multi-channel speech separation (MCSS), including time-frequency (T-F) masking \cite{wang2018multi, chen2019multi, chen2018multi, wang2018spatial}, integration of T-F masking and beamforming \cite{chen2018efficient, drude2017tight}, and end-to-end approaches \cite{gu2019end}. T-F masking based methods formulate speech separation as a supervised learning task in frequency domain. The network learns to estimate a T-F mask for each speaker based on the magnitude spectrogram and interaural differences calculated from the complex spectrograms of observed multi-channel mixture signals, such as the phase difference between two microphone channels, which is known as the interaural phase difference (IPD).

However, one limitation for T-F masking based methods is the phase reconstruction problem. To avoid the complex phase estimation, time-domain speech separation has attracted increasing focus recently. A single-channel time-domain state-of-the-art approach, referred as SC-Conv-TasNet \cite{luo2019convtasnet}, replaces the short time Fourier transform (STFT)-inverse STFT with an encoder-decoder structure. Under the supervision from clean waveforms of speakers, SC-Conv-TasNet's encoder learns to construct an audio representation that optimized for speech separation.
However, the performance of SC-Conv-TasNet is still limited under far-field scenario due to the smearing effects brought by reverberation. To tackle with this problem, in \cite{gu2019end}, we proposed a new MCSS solution, in which hand-crafted IPD features are used to provide spatial characteristic difference information between directional sources. With the aid of additional spatial cues, improved performances have been observed. However, the IPDs are computed in frequency domain with fixed complex filters (i.e., STFT) while the encoder output is learned in the data-driven manner. This causes a data mismatch, which indicates that IPDs may not be the optimal spatial features to incorporate into the end-to-end MCSS framework.

\begin{figure}[t]
\centering
\includegraphics[width=\linewidth]{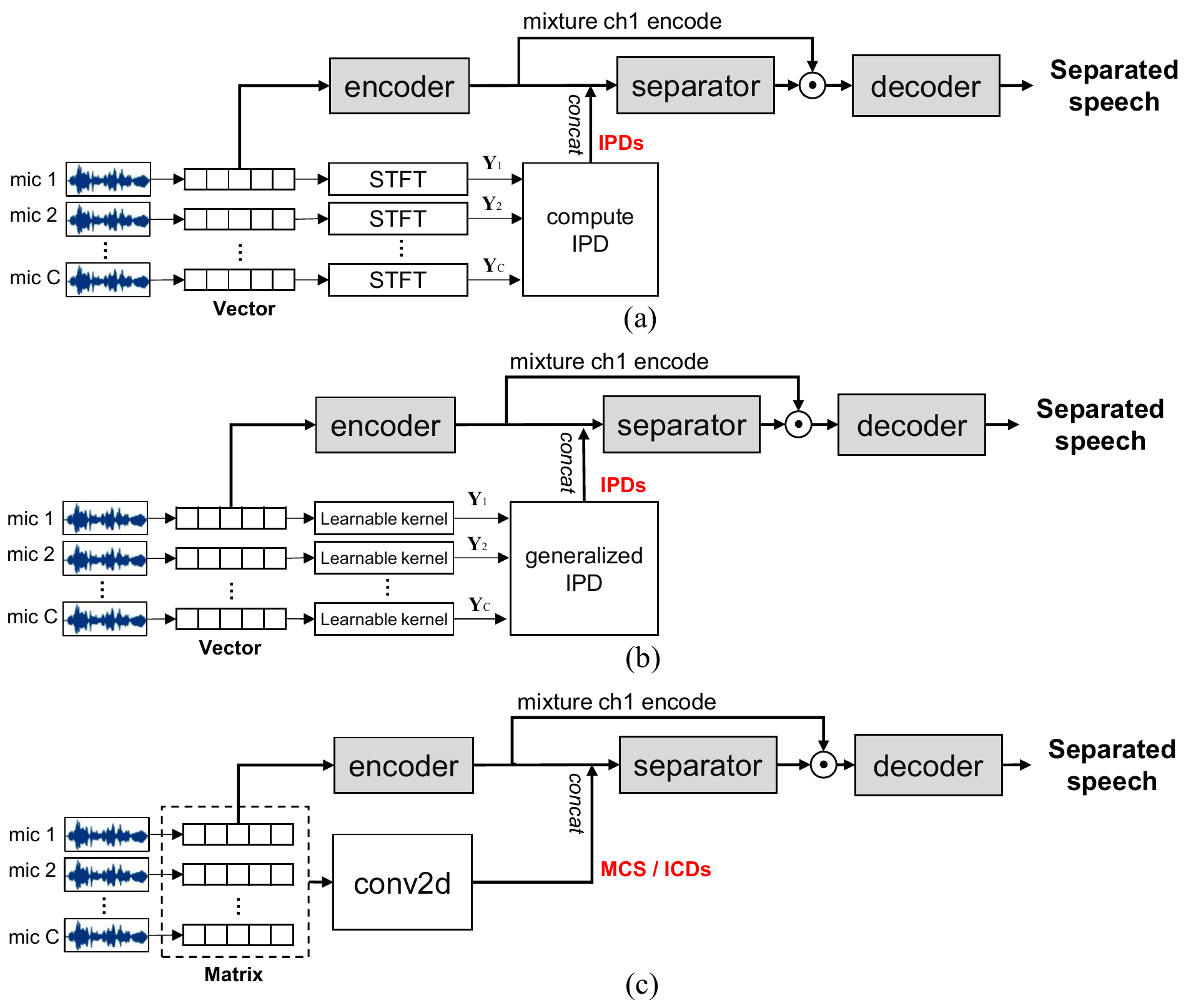}
\caption{The diagram of MCSS model (a) incorporating with IPDs computed by standard STFT. (b) incorporating with generalized IPDs computed by learnable STFT kernel \cite{gu2019end}. (c) incorporating with our proposed MCS and ICDs computed by a conv2d layer.}
\label{fig:framework}
\end{figure}

Bearing above discussions in mind, this work aims to design an end-to-end MCSS model, which are endowed with the capability to learn effective spatial cues using a speech separation objective function in a purely data-driven fashion. As illustrated in Figure \ref{fig:framework} (c), inspired by the success of SC-Conv-TasNet and \cite{gu2019end}, the main body of our proposed MCSS model adopts an encoder-decoder structure. In this design, the time-domain filters spanning all signal channels are trained to perform spatial filtering for multi-channel setting. These filters are implemented by a 2d convolution (conv2d) layer to extract the spatial features. Furthermore, inspired by the formulation of IPD, a novel conv2d kernel is designed to compute the inter-channel convolution differences (ICDs). It is noted that ICDs are learned in data-driven manner and are expected to provide the spatial cues that help to distinguish the directional sources without bringing any data mismatch issue compared with the hand-crafted spatial features. In the end, an end-to-end MCSS model is trained with the scale invariant signal-to-distortion ratio (SI-SDR) loss function. Performance evaluation is conducted on a simulated spatialized WSJ0 2-mix dataset. Experimental results demonstrate that our proposed ICDs based MCSS model outperforms IPD based MCSS model by 10.4\% in terms of SI-SDRi.

The rest of the paper is organized as follows. Section \ref{sec:arch} introduces our proposed architecture in detail. Experimental procedure and result analysis is presented in Section \ref{sec:exp}. Section \ref{sec:conclusion} concludes the paper.

\section{Proposed Architecture}
\label{sec:arch}


\subsection{Multi-channel speech separation}

\begin{figure}[t]
\centering
\includegraphics[width=5cm]{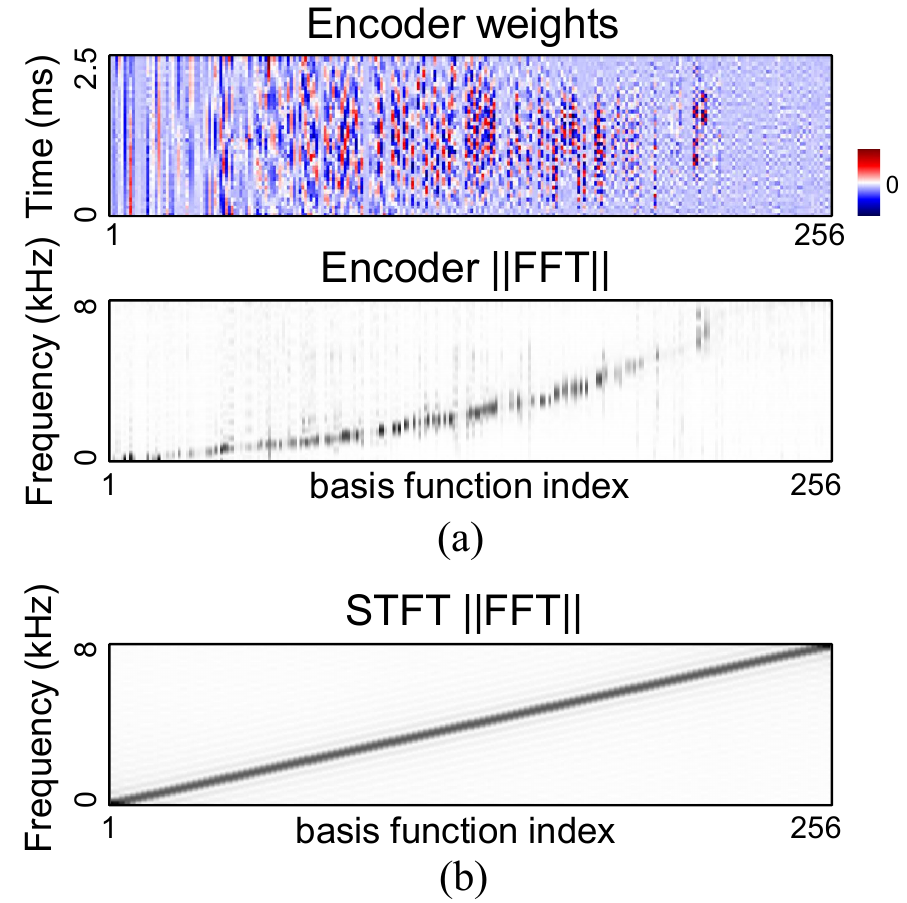}
\caption{(a) Visualization of encoder learned from far-field data. The top figure plots encoder's basis functions and the bottom is the corresponding FFT magnitudes. These basis functions are sorted by the frequency bin containing the peak response. (b) FFT magnitudes of STFT kernel functions.}
\label{fig:fft}
\end{figure}

The baseline MCSS separation system \cite{gu2019end} adopts an encoder-decoder structure, where the data-driven encoder and decoder respectively replaces the STFT and iSTFT operation in existing speech separation pipelines, as shown in Figure \ref{fig:framework}.
Firstly, the encoder transforms each frame of first (reference) channel's mixture waveform $\mathbf{y}_1$ to the mixture encode in a real-valued feature space. Specifically, the learned encoder consists of a set of basis functions, as illustrated in Figure \ref{fig:fft} (a). Most learned filters are tuned to lower frequencies, which shares the similar property with mel filter banks \cite{imai1983cepstral} and frequency distribution of human auditory system \cite{humphries2010tonotopic}. Secondly, IPDs computed by STFT and the mixture encode are concatenated along the feature dimension and fed into the separation module. The separation module learns to estimate a mask in encoder output domain for each speaker, which shares the similar concept with T-F masking based methods. Finally, the decoder reconstructs the separated speech waveform from the masked mixture encode for each speaker. To optimize the network end-to-end, scale-invariant signal-to-distortion ratio (SI-SDR) \cite{le2019sdr} is utilized as the training objective:

\vspace{-0.2cm}
\begin{equation}
\text{SI-SDR}:=10\log_{10}\frac
{\left\|x_{\text{target}}\right\|_{2}^{2}}
{\left\|e_{\text{noise}}\right\|_{2}^{2}}
\label{eq:si_sdr}
\end{equation}
where $x_{\text{target}}:={\left<\hat{x}, x\right>x}/{\left\|x\right\|_{2}^{2}}$, $e_{\text{noise}}:=\hat{x}-x_{\text{target}}$, $x$ and $\hat{x}$ are the reverberant clean and estimated source waveform, respectively. The zero-mean normalization is applied to $x$ and $\hat{x}$ to guarantee the scale invariance.

However, the combination of IPD and encoder output may cause a data mismatch. Different from the encoder which is learned in a data-driven way, the IPD is calculated with complex fixed filters (i.e., STFT), the center frequencies of which are evenly distributed, as illustrated in Figure 1 (b). Also, as \cite{luo2019convtasnet} points out, STFT is a generic transformation for signal analysis that may not be necessarily optimal for speech separation.

\subsection{Spatial feature learning}
\label{sec:spatial_cues}
To perform the spatial feature learning jointly with the rest of the network, we propose to learn spatial features directly from multi-channel waveforms with an integrated architecture. The main idea is to learn time-domain filters spanning all signal channels to perform adaptive spatial filtering \cite{hoshen2015speech, li2016neural, sainath2017raw}. These filters parameters are jointly optimized with the encoder using Eq. \ref{eq:si_sdr} in a purely data-driven fashion.

Denote these filters as $\mathbf{K}=\{\mathbf{k}^{(n)}\} \in \mathbb{R}^{C\times L \times N}$, where $\mathbf{k}^{(n)} = [...,k_c^{(n)}, ...] \in \mathbb{R}^{C\times L}$ is a set of filters spanning $C$ signal channels with window size of $L$. Then, the multi-channel features are computed by summing up the convolution products between the $c$-th channel mixture signal $\mathbf{y}_c$ and filter $k^{(n)}_c$ along signal channel $c$, named as multi-channel convolution sum (MCS):

\vspace{-0.2cm}
\begin{equation}
\text{MCS}^{(n)} = \sum_{c=1}^{C} \mathbf{y}_{c} \circledast k^{(n)}_{c}
    \label{eq:icc}
\end{equation}
where $\circledast$ denotes the convolution operation. The design principle lies in Eq. \ref{eq:icc} is similar to that of delay-and-sum beamformer, where signal arriving at each microphone are summed up with certain time delays to emphasize sound from a particular direction. Each set of filters $\mathbf{k}^{(n)}$ is expected to steer at a different direction, therefore different spatial views of the multi-channel mixture signals can be obtained by MCS and therefore enhancing the separation accuracy.

\begin{figure}[t]
\centering
\includegraphics[width=4.5cm]{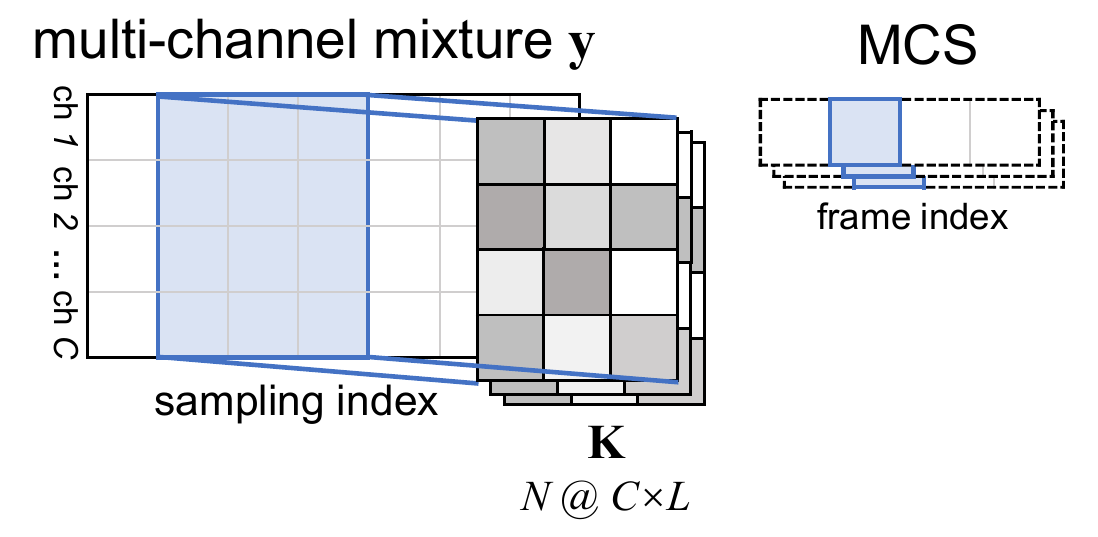}
\caption{Conceptual illustration of the conv2d and generation of different groups of MCS.}
\label{fig:mcs}
\end{figure}

To implement these learnable filters within the network, we employ a 2d convolution (conv2d) layer. The generation of MCS with conv2d is illustrated in Figure \ref{fig:mcs}. The kernel size is $C\times L$ (height$\times$width) and there are $N$ convolution channels in total. The conv2d layer's stride along width axis represents the hop size and is fixed as $L/2$ in our experiments.

Furthermore, inspired by the formulation of interaural differences (e.g., IPDs), we design a special conv2d kernel to extract inter-channel convolution differences (ICDs).
As we know, IPD is a well-established frequency domain feature widely used for spatial clustering algorithms and recent deep learning based MCSS methods. The rationale lies in that, the IPDs of T-F bins that dominated by the same source will naturally form a cluster within each frequency band, since their time delays are approximately the same. The standard IPD is computed by the phase difference between channels of complex spectrogram as $\text{IPD}_{m}=\angle\mathbf{Y}_{m_1}-\angle\mathbf{Y}_{m_2}$, where $\mathbf{Y}$ is the mutli-channel complex spectrogram computed by STFT of multi-channel waveform $\mathbf{y}$, $m_1$ and $m_2$ represent two microphones' indexes of the $m$-th microphone pair.

Following this concept, the $n$-th ICD between the $m$-th pair of signal channels can be computed by:

\begin{figure}[t]
\centering
\includegraphics[width=7cm]{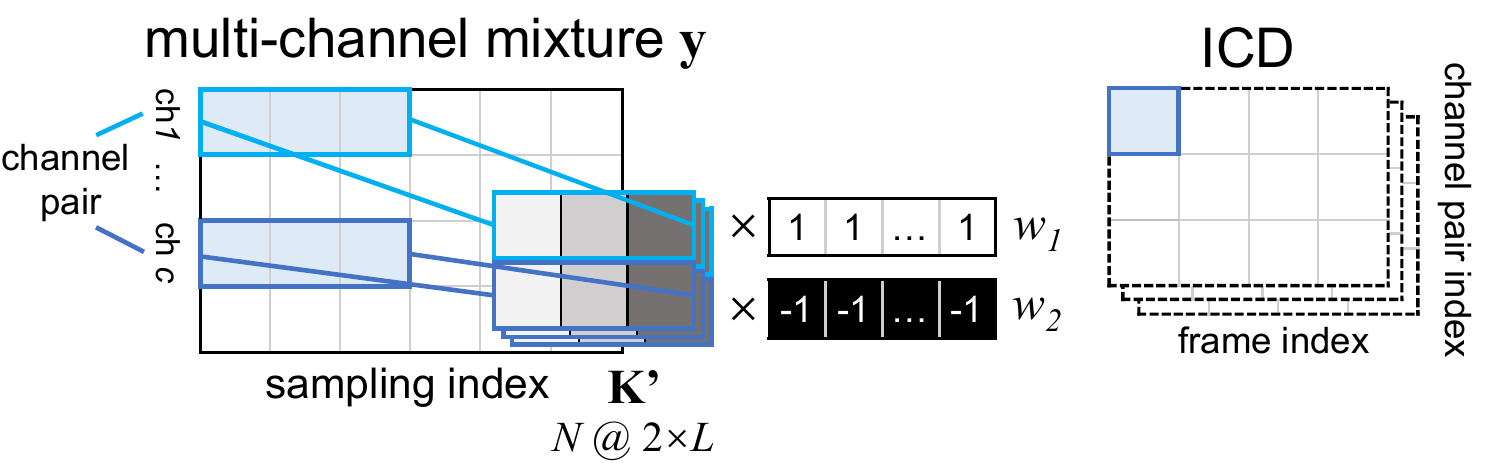}
\caption{Conceptual illustration of our designed conv2d kernel and generation of ICDs. To facinate training, $w_1$ is fixed as 1 while $w_2 \in \mathbb{R}^{1\times L}$ is a learnable parameter initialized as -1. }
\label{fig:icd}
\end{figure}

\vspace{-0.25cm}
\begin{equation}
\text{ICD}^{(n)}_m = \sum_{c=1}^{2} w_c \cdot \left( \mathbf{y}_{m_c} \circledast k'^{(n)} \right)
    \label{eq:icd}
\end{equation}
where $k'^{(n)} \in \mathbb{R}^{1\times L}$ is a filter shared among all signal channels to ensure identical mapping, $w_c \in \mathbb{R}^{1\times L}$ is a window function designed to smooth the ICD and prevent potential spectrum leakage. When $w_1$ is fixed as full ones and $w_2$ as full negative ones, Eq. \ref{eq:icd} calculates the exact inter-channel difference between the $m$-th microphone pair.


Figure \ref{fig:icd} illustrates our designed conv2d kernel and generation of different pairs of ICDs. The conv2d kernel height is set as 2 to span a microphone pair. Note that different configurations of dilation $d$ and stride $s$ on the kernel height axis can extract ICDs from different pairs of signal channels, i.e., $m_1=1+(m-1)s, m_2=2+d+(m-1)s$. For example, for a 6-channel signal, setting dilation as 3 and stride as 1, we can obtain the three pairs of channels: (1, 4), (2, 5) and (3, 6).

\begin{figure}[t]
\centering
\includegraphics[width=5cm]{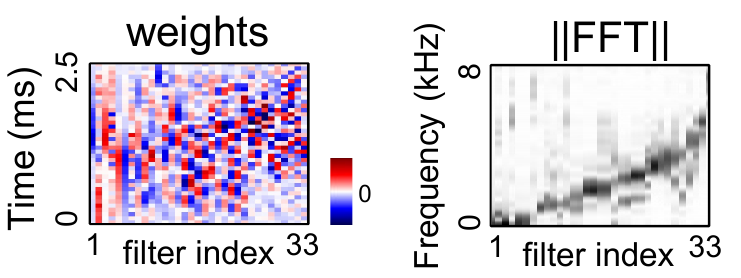}
\caption{Visualization of learned filters $\mathbf{K}'$. Each column represents a filter, and the right plot is its corresponding FFT magnitude response. These filters are sorted by the frequency bin containing the peak response. }
\label{fig:conv2d_kernel}
\end{figure}

To shed light on the property of learned filters $\mathbf{K}'=\{ k'^{(n)}\}$, we visualize these filters in Figure \ref{fig:conv2d_kernel}. It can be observed that these learned filters show similar frequency tuning characteristics with the encoder (Figure \ref{fig:fft} (a)). This suggests that the learned ICD may be more coincident with the encoder output and enables more efficient feature incorporation.

\vspace{-0.2cm}
\section{Experiments and Result Analysis}
\label{sec:exp}

\subsection{Dataset}
We simulated a spatialized reverberant dataset derived from Wall Street Journal 0 (WSJ0) 2-mix corpus, which are open and well-studied datasets used for speech separation \cite{hershey2016deep,yu2017permutation,luo2019convtasnet,wang2019combining}. There are 20,000, 5,000 and 3,000 multi-channel, reverberant, two-speaker mixed speech in training, development and test set respectively. All the data is sampling at 16kHz. The performance evaluation is all done on test set, the speakers in which are all unseen during training. In this study, we take a 6-microphone circular array of 7cm diameter with speakers and the microphone array randomly located in the room. The two speakers and the microphone array are on the same plane and all of them are at least 0.3m away from the wall. The image method \cite{allen1979image} is employed to simulate RIRs randomly from 3000 different room configurations with the size (length-width-height) ranging from 3m-3m-2.5m to 8m-10m-6m. The reverberation time T60 is sampled in a range of 0.05s to 0.5s. Samples with angle difference between two simultaneous speakers of 0-15$\degree$, 15-45$\degree$, 45-90$\degree$ and 90-180$\degree$ respectively account for 16\%, 29\%, 26\% and 29\%.

\begin{table*}[ht]
  \caption{SDRi (dB) and SI-SDRi (dB) performances with different configurations of conv2d layer on far-field WSJ0 2-mix.}
  \label{tab:conv2d}
  \centering
  \begin{tabular}{l|c|c|cccc|c|c}
    \hline
    \multirow{2}{*}{\textbf{Setup}} &
    \multirow{2}{*}{\textbf{window} $w$} &
    \multirow{2}{*}{\textbf{\# filters $N$}} &
    \multicolumn{5}{c|}{\textbf{SI-SDRi (dB)}} &
    \multirow{2}{*}{\textbf{SDRi (dB)}} \\
    & & &$<$15\degree &15\degree-45\degree &45\degree-90\degree &$>$90\degree & Ave. \\
    \hline
    Single-channel Conv-TasNet & -  & - & 8.5 & 9.0  & 9.1	& 9.3	&9.1   & 9.4 \\
    \hline
 +MCS (conv2d (6$\times$40)) & - & 256 &  5.7 & 10.3 & 11.9 & 12.9 & 10.8 & 11.2\\
    \hline
 +ICD (conv2d (2$\times$40)) & fix -1 & 256 & 5.5 & 10.9 & 12.3 & 12.9 & 11.0 & 11.4\\
 +ICD (conv2d (2$\times$40)) & init. -1 & 256 & 6.2 & 11.2 & 12.6 & 13.2 & 11.4 & 11.8\\
    \hline
 +ICD (conv2d (2$\times$40)) & init. randomly & 33 & 8.2 & 8.1 & 9.0 & 9.1 & 8.9 & 9.2 \\
 +ICD (conv2d (2$\times$40)) & fix -1 & 33 & 6.9 & 11.1 & 12.3 & 12.9 & 11.3 & 11.7\\
 +ICD (conv2d (2$\times$40)) & init. -1 & 33 & 6.7 & 11.7 & 13.1 & 13.9 & \textbf{11.9} & \textbf{12.3}\\
    \hline
  \end{tabular}
\end{table*}

\subsection{Network and Training details}

All hyper-parameters are the same with the best setup of Conv-TasNet version 2 in \cite{luo2018surpass}, except $L$ is set to 40 and encoder stride is 20. Batch normalization (BN) is used in all the experiments to speed up the separation process.

The microphone pairs for extracting IPDs and ICDs are (1, 4), (2, 5), (3, 6), (1, 2), (3, 4) and (5, 6) in all experiments. These pairs are selected because the distance of microphones in between each pair is either the furthest or nearest. In this case, there are two setups of dilation $d$ and stride $s$ for the conv2d layer, respectively $d=1,s=2$ and $d=3,s=1$. The first channel of mixture waveform is set as the reference channel as the encoder input. To match the encoder output's time steps, both IPDs and ICDs are extracted with 2.5ms (40-point) window length $L$ and 1.25ms (20-point) hop size with 64 FFT points. SI-SDR (Eq. \ref{eq:si_sdr}) is utilized as training objective. The training uses chunks with 4.0 seconds duration. The batch size is set to 32. Permutation invariant training \cite{yu2017permutation} is adopted to tackle with label permutation problem.

\subsection{Result Analysis}
\label{subsec:rlt}

Following the common speech separation metrics \cite{le2019sdr,vincent2006performance}, we adopt average SI-SDR and SDR improvement over mixture as the evaluation metrics. We also report the performances under different ranges of angle difference between speakers to give a more comprehensive assessment for the model.

\noindent\textbf{Different configurations for conv2d layer}. We explore different conv2d configurations for computing the ICD, including different numbers of filters and initialization methods of window function $w$ ($w_2$ in section \ref{sec:spatial_cues}). The number of filters are chosen to be 256 and 33, where 256 matches the basis function number of encoder, 33 is the number of bins for 64-point FFT size, which is the closest exponential of 2 for 40-point frame length. The results are listed in Table \ref{tab:conv2d}. SC-Conv-TasNet is served as the baseline system, achieving 9.1dB of SI-SDRi on the far-field dataset. By learning spatial filters, the MCS based model outperforms the baseline by 1.7dB of SI-SDRi. For ICD setups, we found that the performances with 33 filters are relatively superior to those with 256 filters. One possible reason is that, according to sampling theorem, the highest frequency resolution can achieve with sampling rate of 16kHz and frame length of 40 is limited.

Furthermore, the value of $w$ contributes significantly to the separation performance (9.2dB v.s. 12.3dB for model with 33 filters). If $w$ is randomly initialized, or in other words, there is no explicit subtraction operation between signal channels, the model will not be able to automatically learn useful spatial cues. If $w$ is initialized and fixed as -1 (fix -1), this indicates that the exact convolution difference operation between signal channels is computed as the ICD. Furthermore, relaxing $w$ to be learnable (init. -1) produces a much better result, which demonstrates the validity of ICD's formulation.

\begin{table}[t]
  \caption{SDRi (dB) and SI-SDRi (dB) performances of IPD, ICD-based separation systems on far-field WSJ0 2-mix.}
  \label{tab:ipd_vs_pipd}
  \centering
  \begin{tabular}{p{150 pt}|p{14 pt}<{\centering}p{18 pt}<{\centering}|c}
    \hline
    \multirow{2}{*}{\textbf{Features}} &
    \multicolumn{3}{c}{\textbf{SI-SDRi (dB)}}\\
    & $<$15\degree &$>$15\degree & Ave. \\
    \hline
 cosIPD, sinIPD  & 7.7 & 12.2 & 11.5\\
 cosIPD, sinIPD (learnable kernel) \cite{gu2019end}  & 7.9 & 12.3 & 11.6  \\
    \hline
 ICD  & 6.7 & 12.9 & 11.9 \\
 ICD, cosIPD, sinIPD & 8.1 & 13.2 & \textbf{12.4} \\
    \hline
  \end{tabular}
\end{table}

\begin{figure}[t]
\centering
\includegraphics[width=6cm]{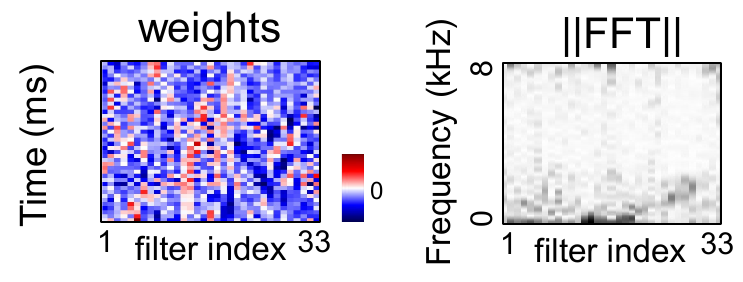}
\caption{Visualization of learned filters when incorporating ICD with cosIPD and sinIPD. Each column represents a filter, and the right plot is its corresponding FFT magnitude response. These filters are sorted by the frequency bin containing the peak response.}
\label{fig:conv2d+ipd_kernel}
\end{figure}

\noindent\textbf{IPD versus ICD}. We examine the performance of IPD versus proposed ICD for MCSS and report the results in Table \ref{tab:ipd_vs_pipd}. In addition, the performance of IPD with trainable kernel based MCSS model \cite{gu2019end} is listed for comparison. Specifically, in \cite{gu2019end}, the standard STFT operation is reformulated as a function of time domain convolution with a trainable kernel, which is optimized for the speech separation task.
Combining the cosIPD and sinIPD we can obtain SI-SDRi of 11.5dB, which has 2.4dB gain over the single-channel baseline. It suggests IPDs can provide beneficial spatial information of sources. With the trainable kernel, the performance improves slightly.
The proposed ICD based separation model obtains 0.4dB improvement over cosIPD+sinIPD based, benefiting from the data-driven learning fashion. Note that the performance under 15$\degree$ for ICD based model is worse than that of IPD based. One possible reason is that the portion of data under 15$\degree$ is relatively few hence causing difficulty in learning effective ICDs. The incorporation of ICDs and IPDs achieves further 0.5dB improvement. In this case, we also visualize the learned filters in Figure \ref{fig:conv2d+ipd_kernel}, which show different patterns from those in Figure \ref{fig:conv2d_kernel}. We found that almost all filters are tuned to relatively low frequency. This indicates that the ICDs may learn complementary spatial information to compensate the IPD ambiguity in low frequencies.

\section{Conclusion}
\label{sec:conclusion}
This work proposes an end-to-end multi-channel speech separation model, which is able to learn effective spatial cues directly from the multi-channel speech waveforms in a purely data-driven fashion. Experimental results demonstrated the MCSS model based on learned ICDs outperforms that based on well established IPDs.

\vfill\pagebreak

\bibliographystyle{IEEEtran}
\bibliography{mybib}

\end{document}